\documentclass[aps,prd,showpacs,twocolumn]{revtex4}

\usepackage{epsfig}
\usepackage{longtable}

\begin{document}

\title{Study of $V_LV_L\to t\bar{t}$ at the ILC Including 
${\cal O} (\alpha_s) $ QCD Corrections} 
\author{Stephen Godfrey$^a$\footnote{Email: godfrey@physics.carleton.ca}}
\author{Shou-hua Zhu$^{a,b}$}
\affiliation{
$^a$Ottawa-Carleton Institute for Physics, Department of Physics, \\
Carleton University, Ottawa, Canada K1S 5B6\\
$^b$Institute of Theoretical Physics, School of Physics, \\
Peking University, Beijing 100871, China}

\date{\today}

\begin{abstract}
In the event that the Higgs mass is large or that the electroweak 
interactions are strongly interacting at high energy, top quark 
couplings to longitudinal components of the weak gauge bosons could 
offer important clues to the underlying dynamics.  
It has been suggested that 
precision measurements of $W_L W_L \to t\bar{t}$ and $Z_L Z_L \to t\bar{t}$ 
might provide hints of new physics.  In this paper we present results
for ${\cal O} (\alpha_s) $ QCD corrections to
$V_LV_L\to t\bar{t}$ scattering at the ILC.  We find that corrections to 
cross sections  can be as large as 30\% and must be 
accounted for in any precision measurement of $VV\to t\bar{t}$.
\end{abstract}

\maketitle

\section{Introduction}

Understanding the mechanism of electroweak symmetry breaking (EWSB) is a 
primary goal of the Large Hadron Collider (LHC) \cite{Barbieri:2003dd}.  
Numerous 
possibilities have been proposed and in some cases their experimental 
signatures have been studied in detail \cite{atlas:1999fr}.  
The models can
be roughly divided into two scenarios.  
The first possibility is a weakly interacting weak sector which 
implies light Higgs bosons and naturally leads to Supersymmetry which 
predicts many new supersymmetric particles \cite{Baer:1995tb}.  
The alternative scenario is a strongly interacting Higgs sector (SIWS)
\cite{Barklow:1997nf}
most often described by dynamical symmetry breaking \cite{Hill:2002ap}.  
Recently, additional new 
approaches to EWSB  have emerged, such as the Little Higgs model 
\cite{Arkani-Hamed:2001nc,Low:2002ws,Schmaltz:2002wx,Han:2003wu,Hewett:2002px,Perelstein:2003wd}
which contain ingredients from both scenarios
and other alternative approaches such as ``Higgsless'' models 
\cite{Csaki:2003dt,Csaki:2003zu,Davoudiasl:2003me}. 
Although the light Higgs mass scenario is favoured by precision 
electroweak fits, models with heavy Higgs bosons can be accomodated by 
the data \cite{Peskin:2001rw,Chanowitz:2002cd,Barbieri:2004qk}. It is this
latter possibility that we wish to study.

While the signature for the weakly interacting weak sector is the 
production of light Higgs bosons and possibly additional particles and many
of the newer models predict relatively light extra gauge bosons, the 
signal for a SIWS will almost certainly be more subtle \cite{Group:2004hn}.  
In the SIWS 
scenario $W$-boson interactions can be described by an effective 
Lagrangian with the coefficients in $L_{eff}$ constrained by 
experiment \cite{Dawson:1990cc}.  
Detailed studies have been made for the LHC 
\cite{Chanowitz:1998wi,Rosenfeld:1988th,Bagger:1993zf,Barklow:1997nf}.  

Because the $t$-quark mass is the same order of magnitude as the scale of 
EWSB it has long been suspected that $t$-quark properties may provide 
hints about the nature of EWSB 
\cite{Chanowitz:1978mv,Larios:1996ib,Zhang:1994fb,Whisnant:1994fh,Wudka:1997sk,Han:2000ic,Han:2004zh}.  
To this end studies have been 
performed of $VV\to t\bar{t}$ both in terms of the sensitivity to 
$M_H$ \cite{Gintner:1996cr,RuizMorales:1999kz,Alcaraz:2000xr}, 
and in terms of an effective Lagrangian describing vector boson 
$t$-quark interactions \cite{Larios:2000xj}.  
Moreover, this process 
should be sensitive to models such as top-colour \cite{Hill:1991at}
and the top see-saw model \cite{Dobrescu:1997nm}.
Due to the Goldstone boson equivalence theorem \cite{et} the 
longitudinal gauge bosons $(W^\pm_L, \; Z_L)$ are equivalent to the 
Goldstone modes at energies much larger than their mass so that they 
reflect the properties of EWSB.
In principle, $V_LV_L\to t\bar{t}$ can be studied at both hadron 
colliders through $q_1 q_2 \to q_1' q_2' V_L V_L'$ and at high energy 
$e^+e^-$ colliders via $e^+e^-\to \ell_1 \ell_2 V_L V_L'$ or 
$\gamma\gamma \to  V_L V_L' +X $ 
\cite{Kauffman:1989aq,Larios:1996ib,Gintner:1996cr,RuizMorales:1999kz,Alcaraz:2000xr}.  
While the LHC will be the first 
high energy collider to explore the TeV energy region the overwhelming 
QCD backgrounds for $t\bar{t}$ production will 
likely make it impossible to study 
the $V_LV_L\to t\bar{t}$ subprocess \cite{Han:2003pu}.  

The ILC offers a 
much cleaner environment to study the $V_LV_L\to t\bar{t}$ subprocess.
The simplist 
approach is to study how the $t\bar{t}$ cross section varies with 
$M_H$.  This has been studied by several authors 
\cite{Gintner:1996cr,RuizMorales:1999kz,Alcaraz:2000xr}. 
The tree-level Feynman diagrams for the SM process are given in Fig.~1.
The $e^+e^-\to \nu \bar{\nu} t\bar{t}$ cross 
section is shown as a function of $M_H$ in Fig.~2
for an $e^+e^-$ collider with centre of mass energy 
$\sqrt{s}_{e^+e^-}=1$~TeV. 
We convoluted the effective $W$ 
approximation (EWA) distributions \cite{eva,Dawson:1986tc}
with the $W^+_LW^-_L \to t\bar{t}$ cross 
section to obtain the $e^+e^-\to \nu \bar{\nu} t\bar{t}$ cross 
sections and included the kinematic cuts 
$M_{t\bar{t}}>400$~GeV and $p_T^{t,\bar{t}}>10$~GeV.  The values for
these kinematic cuts were chosen so that in the former case the EWA 
is reasonably reliable and in the latter case to reduce various 
backgrounds.  This is discussed more fully in section IV.

\begin{figure}[b]
\begin{center}
\centerline{\epsfig{file=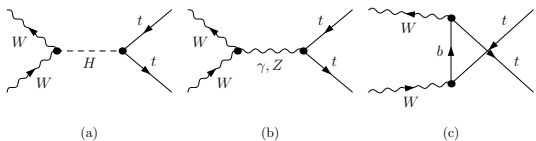,width=2.8in,clip=}}
\end{center}
\caption{The tree level diagrams for $W^+W^-\to t\bar{t}$. }
\end{figure}

\begin{figure}[h]
\begin{center}
\centerline{\epsfig{file=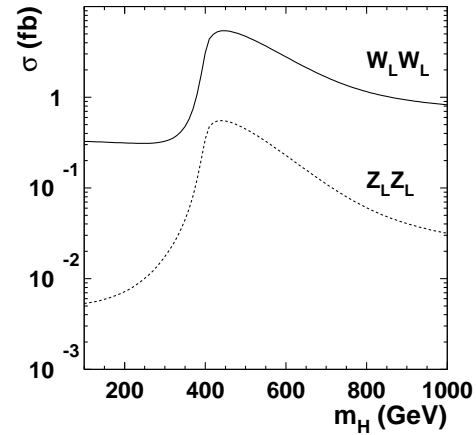,width=2.5in,clip=}}
\end{center}
\caption{The solid line is for
$\sigma(e^+e^- \to \nu \bar{\nu} t\bar{t})$ (the $W^+_LW^-_L\to t\bar{t}$ 
subprocess) and the dotted line is for 
$\sigma(e^+e^- \to e^+ e^- t\bar{t})$ (the $Z_LZ_L\to t\bar{t}$ 
subprocess) as a function of 
$M_H$ for an $e^+e^-$ collider with centre of mass energy 
$\sqrt{s}_{e^+e^-}=1$~TeV and kinematic cuts  
$m_{t\bar{t}}>400$~GeV and $p_T^{t,\bar{t}}>10$~GeV.}
\end{figure}

A more general
approach is to parametrize interactions in a nonlinearly realized 
electroweak chiral Lagrangian \cite{Dawson:1990cc} which
is appropriate if the EWSB dynamics is 
strong with no Higgs bosons at low energies.  For illustrative 
purposes we show the sensitivity of the cross sections 
to a representitive dimension five operator;
\begin{equation}
L^{eff}_1= {{a_1}\over{\Lambda}} \bar{t} t W_\mu^+ W^{-\mu}
\end{equation}
where the coefficient $a_1$ is naively expected to be of order 1 when 
the cut-off of the theory is taken to be $\Lambda=4\pi v=3.1$~TeV 
with $v=246$~GeV, the vacuum expectation value of the SM Higgs field. 
We do not rigorously follow the chiral Lagrangian approach and simply 
include the operator of Eqn. 1 as an additional $WWt\bar{t}$ 
interaction.  
The effect of varying $a_1$ for an $e^+e^-$ collider with 
$\sqrt{s}_{e^+e^-}=1$~TeV is shown in Fig.~3  
using the EWA and the same kinematic cuts as before.
$\sigma(e^+e^-\to \nu \bar{\nu} t\bar{t})$ is shown for $M_H=120$, 
500, and 1000~GeV and for $M_H=\infty$.  The latter case is 
referred to in the 
literature as the ``low energy theorem'' (LET) model. 
A careful analysis by
Ref.~\cite{Larios:1997ey,Larios:2000xj} 
finds that the  $V_LV_L t\bar{t}$ coupling can be measured to an 
accuracy of $\sim \pm 0.1$ at 95\% C.L. 
(with the cut-off scale of $\Lambda=3.1$~TeV divided out as expressed 
in Eqn. 1).

\begin{figure}[t]
\begin{center}
\centerline{\epsfig{file=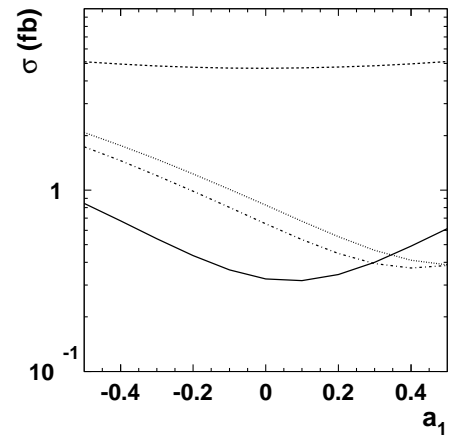,width=2.5in,clip=}}
\end{center}
\caption{$\sigma(e^+e^- \to \nu \bar{\nu} t\bar{t})$ 
(the $W^+_LW^-_L\to t\bar{t}$ subprocess) vs the $WW t\bar{t}$ coupling 
$a_1$ for $\sqrt{s}_{e^+e^-}=1$~TeV with the same kinematic cuts as 
Fig.~2.  The solid line is for $M_H=120$~GeV, 
the dashed line for $M_H=500$~GeV,
the dotted line for $M_H=1$~TeV, and the
the dot-dashed line for $M_H=\infty$ (the LET scenario). }
\end{figure}

These approaches are complementary to studies based on specific 
models and on the linear realization of EW symmetry breaking which 
includes physical Higgs bosons in the particle spectrum.  The crucial
point is that precision measurements of top-quark interactions could 
play an 
important role in understanding the mechanism of EWSB.  But to
be able to attach meaning to precision 
measurements it is necessary to understand radiative corrections, both 
electroweak and QCD.  

In this paper we quantify the importance of QCD 
radiative corrections for the process $V_L V_L \to t\bar{t}$. 
We focus on the ${\cal O} (\alpha_s) $ QCD
corrections to the tree level electroweak 
$V_L V_L \rightarrow t \bar t$ process in the SM at the ILC.
We begin in section II by summarizing the effective $W$ approximation
and how it is implemented in our calculations. The bulk of the 
section is devoted to how we calculated the QCD corrections.  
Numerical results are given in section III with section IV devoted to a 
discussion of other aspects which should be considered in a 
complete calculation of $t\bar{t}$ production. Concluding comments are 
given in the final section.

\section{Calculations}

We are interested in the subprocesses $VV\to t\bar{t}$ which occur in 
the processes
\begin{equation}
e^+e^- \to  \ell_1 \ell_2 +  V V \to \ell_1 \ell_2 + t\bar{t}
\end{equation}
where $\ell_1 \ell_2$ is $\nu\bar{\nu}$ for the $W^+W^-\to t\bar{t}$ 
subprocess and $e^+e^-$ for the $ZZ \to t\bar{t}$ subprocess. Before 
proceeding to the QCD corrections we briefly summarize the 
effective vector boson approximation (EVA).

\subsection{The Effective Vector Boson Approximation}

In the effective vector boson approximation the $W$ and $Z$ 
bosons are treated as 
partons inside the electron \cite{eva,Dawson:1986tc}.  
The total cross section is then
obtained by integrating the $W$ (of $Z$) 
luminosities with the subprocess cross section \cite{Kauffman:1989aq}.
\begin{equation}
\sigma(e^+e^-\to \ell_1 \ell_2 t\bar{t})
= \sum_{\lambda_1, \lambda_2} \int_{\hat{s}/m_{t\bar{t}}^2}^1 \; d\tau \; 
{{d{\cal L}}\over {d\tau}} \; \hat{\sigma}(V_{\lambda_1}V_{\lambda_2} 
\to t\bar{t})
\end{equation}
where $\lambda_1$ and $\lambda_2$ are the $V$'s polarization, 
$\tau=\hat{s}/s$ with $s$ and $\hat{s}$ the $e^+e^-$ and $VV$
 centre-of-mass energy respectively, and 
${\cal L}$ is the $V_{\lambda_1}V_{\lambda_2}$ luminosity 
given by 
\begin{equation}
{{d{\cal L}}\over {d\tau}} = \int_\tau^1 {{dx}\over x} 
f_{V_{\lambda_1}/e}(x)  f_{V_{\lambda_2}/e} (\tau/x).
\end{equation}
The $f_{V/e}$ distributions are given by \cite{eva,Dawson:1986tc}
\begin{eqnarray}
f_{V_\pm/e}(x) & = &{{C}\over{16\pi^2 x}} \left[{ (g_V^f\mp g_A^f)^2  
} \right.
\nonumber \\
& &  \left. { +(g_V^f\pm g_A^f)^2 \; (1-x)^2 }\right] \; 
\log \left( { {4E^2}\over{M_V^2} } \right) \\
f_{V_L/e}(x) & = &C {{(g_V^f)^2 + (g_A^f)^2 }\over{4\pi^2 }}
\left[ { {1-x}\over x} \right]
\end{eqnarray}
for the transverse and longitudinal $V$'s respectively and 
where for a $W$ boson $C=g^2/8$, $g_V=-g_A=1$ and for a $Z$ boson
$C=g^2/\cos^2\theta_w$, 
$g_V=\frac{1}{2} T_3 - Q \; \sin^2 \theta_w$, $g_A = -\frac{1}{2} T_3 $
with $g$ the coupling constant for the weak-isospin group $SU(2)_L$
and $E$ the energy of the initial lepton.   
With these expressions for the $V$ distributions, the luminosity 
for $LL$ scattering simplifies to \cite{eva,Dawson:1986tc}:
\begin{equation}
{ {d{\cal L}_{LL}}\over {d\tau}} = 
{{[(g_V^e)^2 + (g_A^e)^2]^2 }\over {16 \pi^4 \tau}} [ 2(\tau-1) -(1+\tau)\log \tau ]
\end{equation}

\subsection{${\cal O}(\alpha_s)$ corrections to $VV\to t\bar{t}$}

We next describe how we calculated the ${\cal O}(\alpha_s)$ corrections for 
the process $W^+W^-\to t\bar{t}$.  The results for $ZZ\to t\bar{t}$ 
are obtained the same way.  To obtain our results we used the 
FeynArts, FormCalc and LoopTools packages \cite{fclt}.  We 
point out details specific to our calculation but refer the 
interested reader to the descriptions of the packages for further 
information.

The tree-level 
Feynman diagrams for $W^+W^- \to t\bar{t}$ are given in Fig. 1 and 
the virtual diagrams are shown in Fig. 4a. 
The infrared singularity in the vertex 
corrections are cancelled by the soft contributions from the process 
$W^+W^-\to t\bar{t}g$ which are shown in Fig. 4b. 
Since only soft photon formulas are embedded in the 
FeynArts/ FormCalc/ LoopTools packages, to accomodate the soft gluons 
we modified the corresponding formulas with the substitution $eQ\to g_s T_a$.
For processes with no triple gluon vertex present, introducing a gluon 
mass is equivalent to standard dimensional regularization.  
Introducing a small finite gluon mass is easily included using the 
packages we employ, so for convenience we regulate the IR-singularity this way.
This approach has the additional benefit that
varying the value of the gluon mass acts as a check of 
the numerical cancellations between the different contributions.

\begin{figure}[t]
\begin{center}
\centerline{\epsfig{file=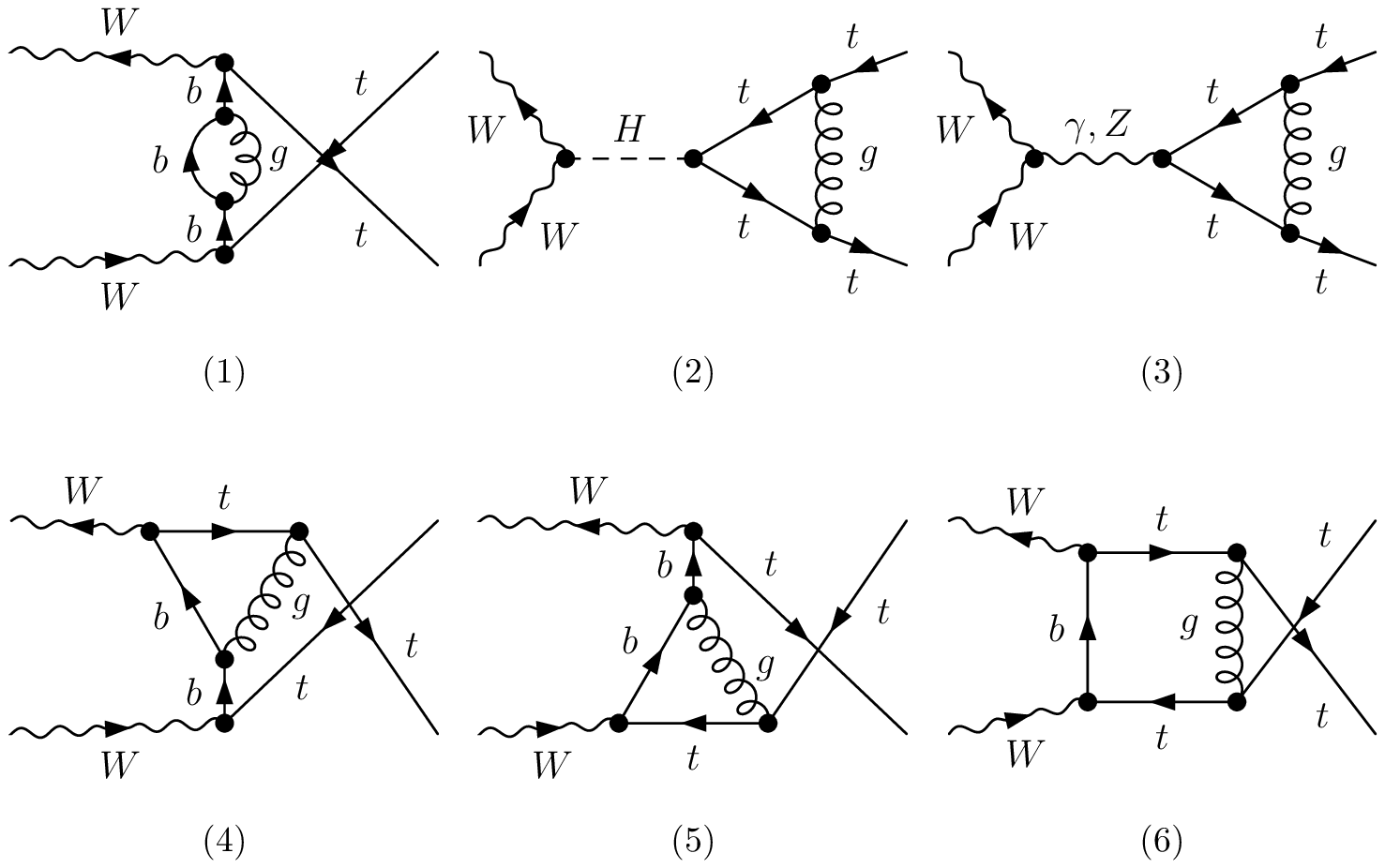,width=2.8in,clip=}}
\end{center}
\centerline{(a)}
\begin{center}
\centerline{\epsfig{file=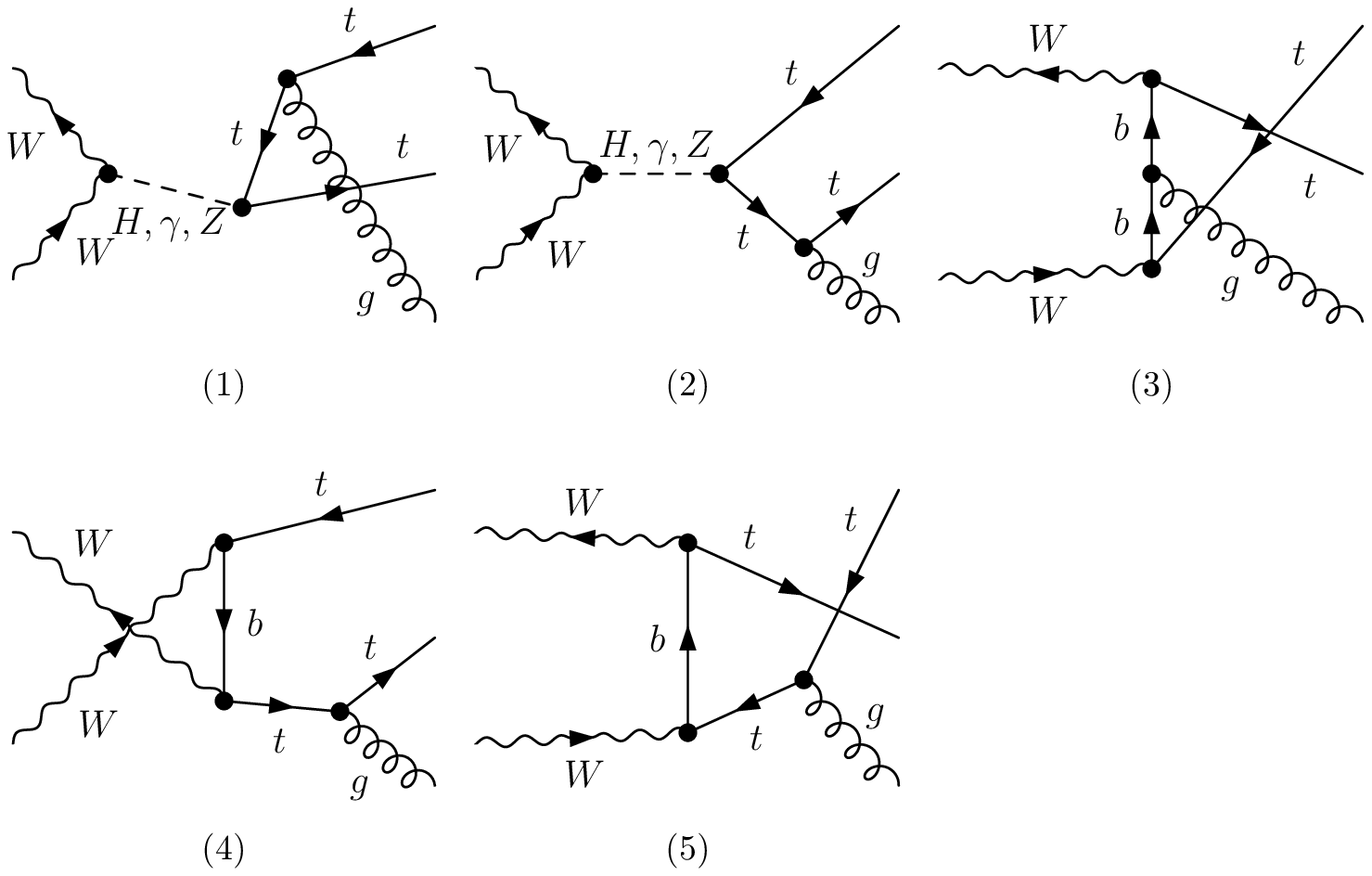,width=2.8in,clip=}}
\end{center}
\centerline{(b)}
\caption{${\cal O} (\alpha_s) $ QCD corrections to $W^+W^-\to t\bar{t}$.
(a) Virtual QCD contributions to $W^+W^-\to t\bar{t}$.
(b) Feynman diagrams for $W^+W^-\to t\bar{t}+g$. }
\end{figure}

The cross-sections are calculated by replacing the Born matrix 
element squared by 
\begin{equation}
|M_{Born}|^2 \to |M_{Born}|^2(1+\delta_{soft}) +2 {\rm Re}(M_{Born}^*\delta M)
\end{equation}
where $\delta M$ is the sum of the one-loop Feynman diagrams 
and the corresponding counter-term diagrams and 
$\delta_{soft}$ is the soft-gluon correction factor coming from the 
$2\to 3$ process. The hard contribution from the $2\to 3$ process is 
also ${\cal O}(\alpha_s)$ and will be discussed further below.
Thus, Eqn.~8 yields the 
cross-section including all ${\cal O}(\alpha_s)$ corrections but 
neglects the ${\cal O}(\alpha_s^2)$ corrections that are included in 
the dropped $|\delta M|^2$ contribution.

To deal with renormalization associated with the ultraviolet 
divergences we adopt the on-mass shell renormalization scheme and use
dimensional regularization.  
In the on-mass shell renormalization scheme for the external quark 
legs, the top-quark self-energy is cancelled by its corresponding 
counter-term.  For this reason we do not explicitely 
show the top-quark self energy diagrams in Fig. 4.
The scale dependence in this calculation only enters in $\alpha_s$ and 
we take it to be $\hat{s}$. We will discuss uncertainties due to this 
choice below.

The software packages we used handle renormalization by
employing numerical 
factors.  For example, the UV divergence is represented by a numerical 
factor $\Delta$.
Because our results must be independent of these factors 
we can vary them to check the
consistency of our results. In particular, the UV 
finiteness is checked by verifying the independence of the results to 
$\Delta$ and in the on-mass shell renormalization scheme to the 
dimensional regularization scale parameter $\mu$. 

As mentioned above, the IR-singularity was regulated by introducing a 
finite gluon mass which is allowed when no triple-gluon vertex 
contributes to the amplitude.  The cancellation of the finite gluon 
contributions from the vertex correction with the soft gluon 
production in $W^+W^-\to t\bar{t}g$ is a strong test of the veracity 
of our results.  We performed this check and also verified that 
the results are independent of the gluon 
regulator mass but for the sake of brevity do not show plots of these 
results.

A final check is to verify that the results are independent of the soft 
cutoff energy of the emitted gluon which divides the cross section 
into a piece with a soft gluon emitted and a piece with a hard gluon 
emitted.  To do so we sum the results of the hard $WW\to t\bar{t}g$ 
process with the soft $WW\to t\bar{t}g$ plus 1-loop results.  While 
individually the hard and soft pieces are dependent on the soft 
cutoff energy of the emitted gluon, the sum will not be.  We have 
checked that our results 
are indeed independent of the gluon cutoff energy.

To obtain cross sections from the analytical results for the squared 
amplitudes, the FormCalc package
integrates phase space using gaussian quadrature for 
the 2 to 2 process and the VEGAS Monte Carlo integration package 
for the 2 to 3 process.

\section{Results}

\begin{figure}[t]
\centerline{\epsfig{file=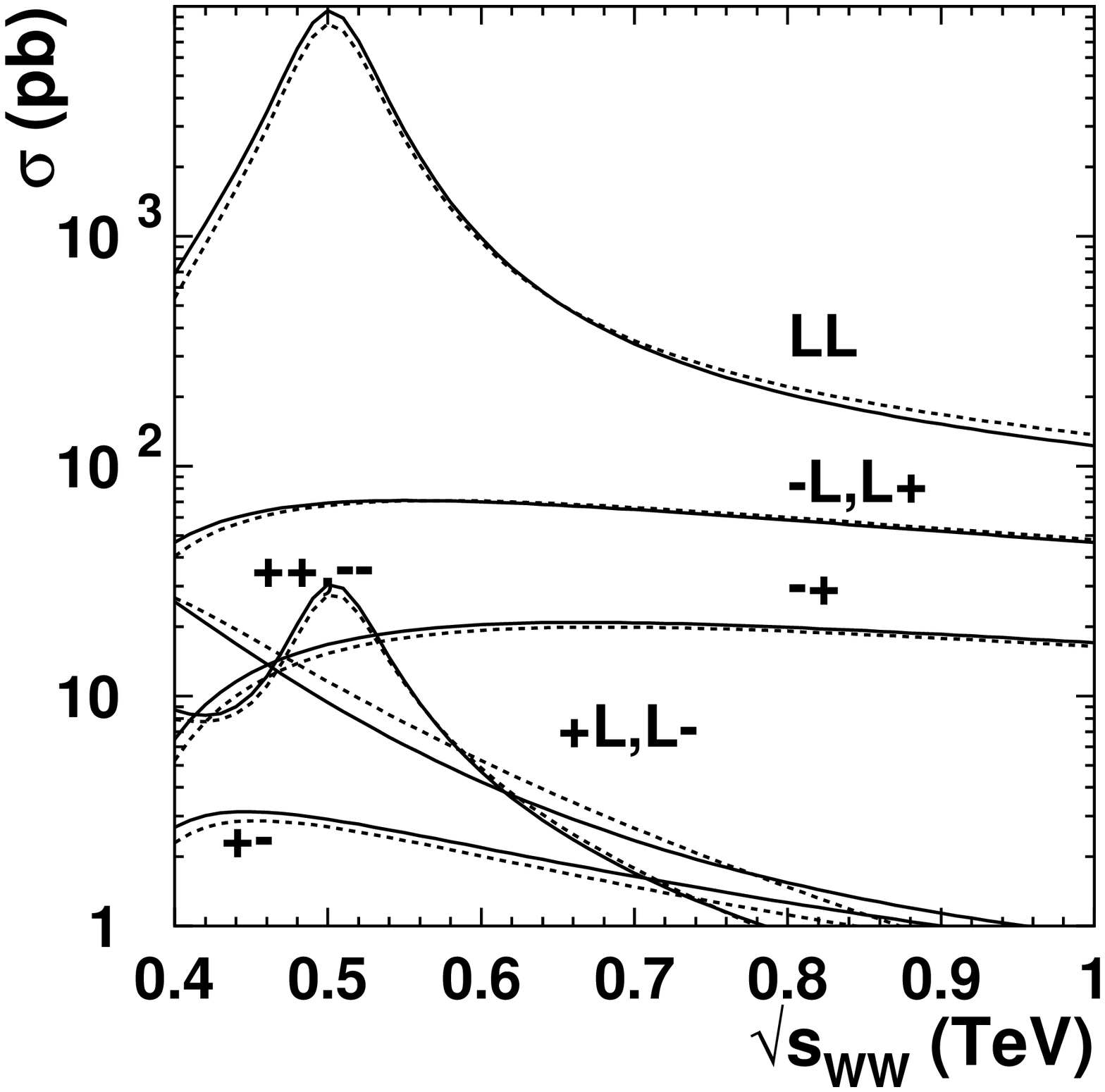,width=2.5in,clip=}}
\centerline{\epsfig{file=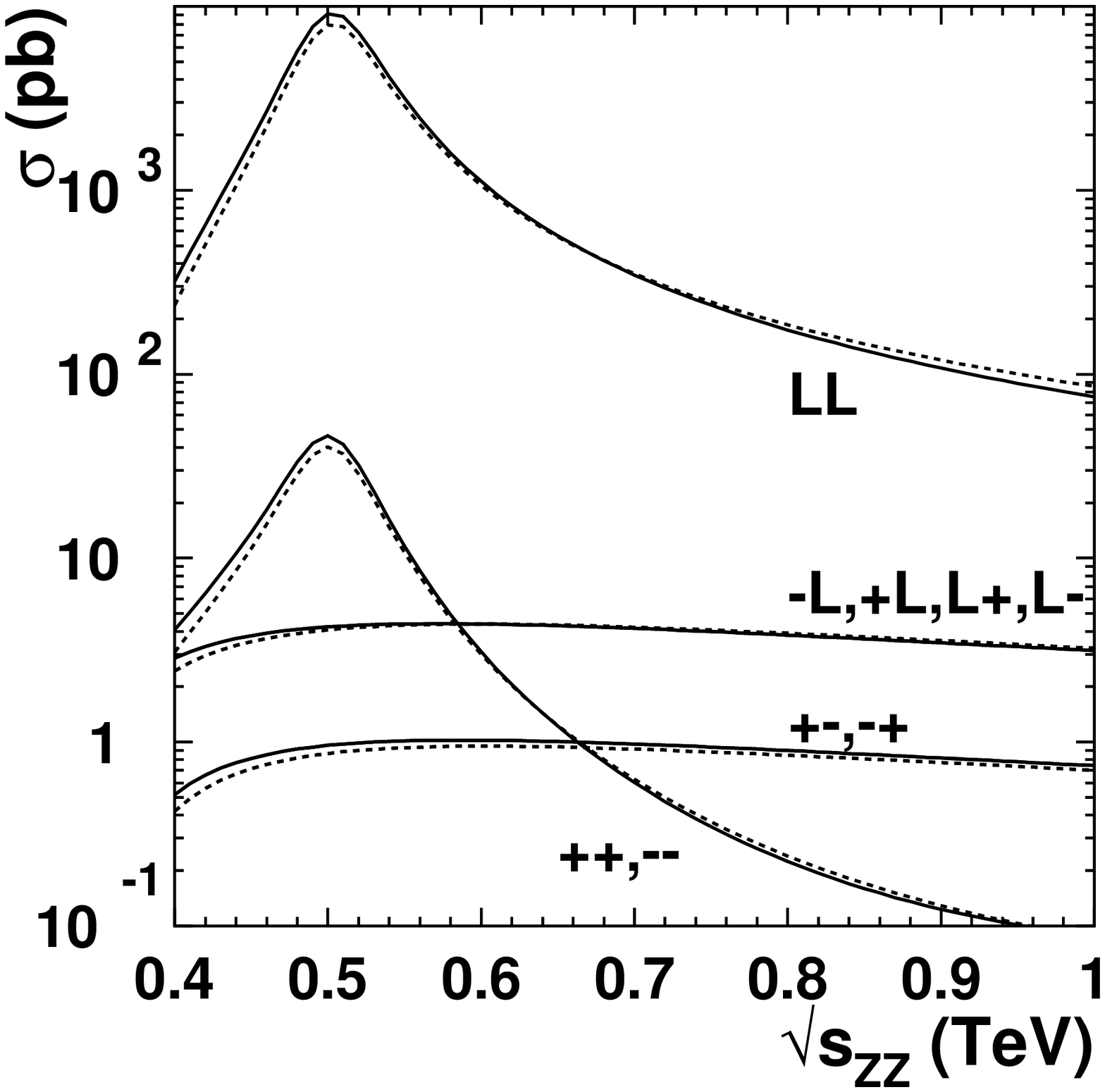,width=2.5in,clip=}}
\caption[]{
Cross sections as a function of $\sqrt{\hat{s}}$ for the subprocesses 
(a) $W^+W^- \rightarrow t \bar t$ 
and (b) $ZZ \rightarrow t \bar t$.  In both cases 
$m_H=500$ GeV for the ${\cal O} (\alpha_s) $ 
QCD corrected (solid) and electroweak tree level (dashed). } \label{ww}
\end{figure}

\begin{figure}[t]
\centerline{\epsfig{file=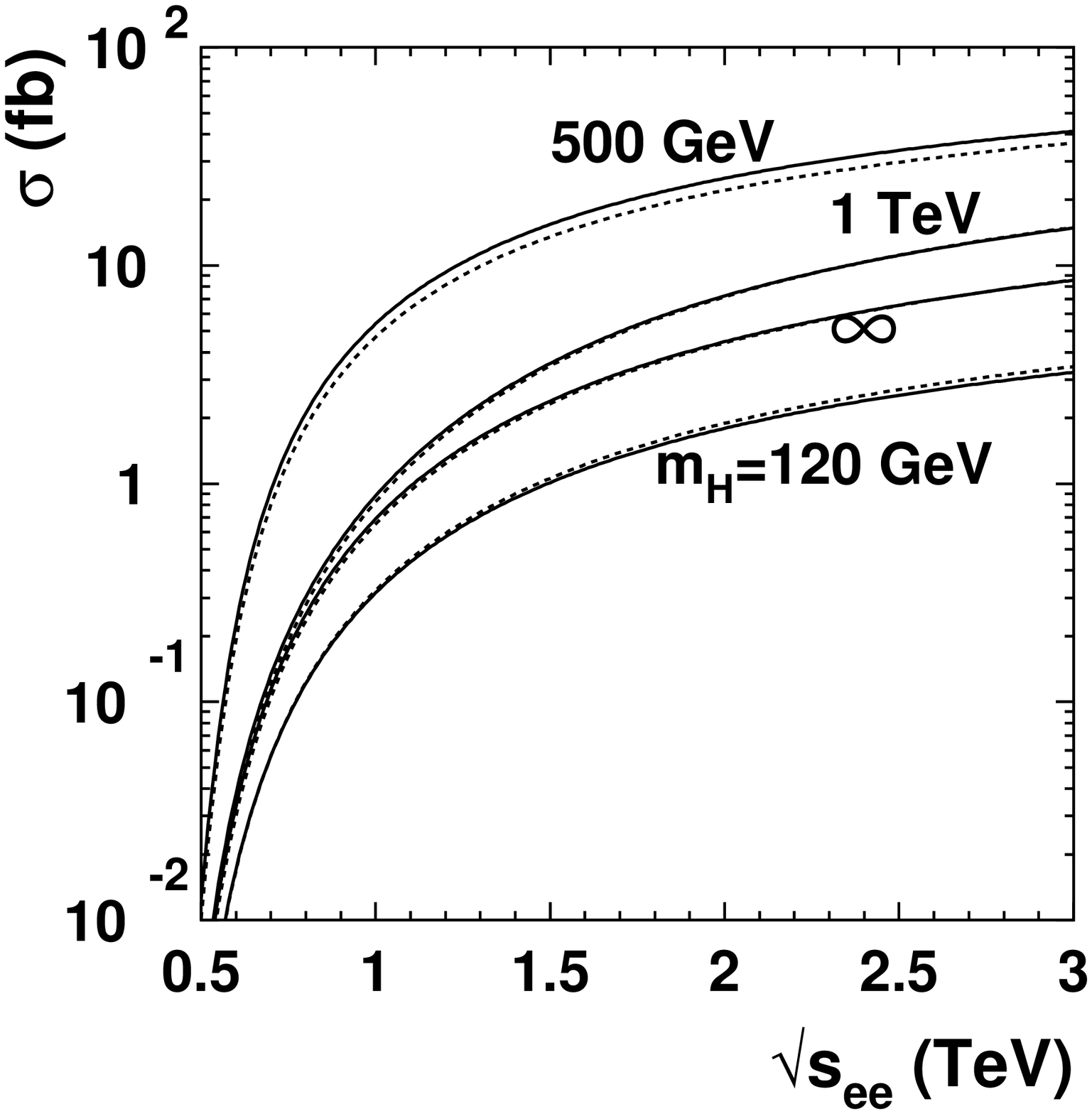,width=2.5in,clip=}}
\centerline{\epsfig{file=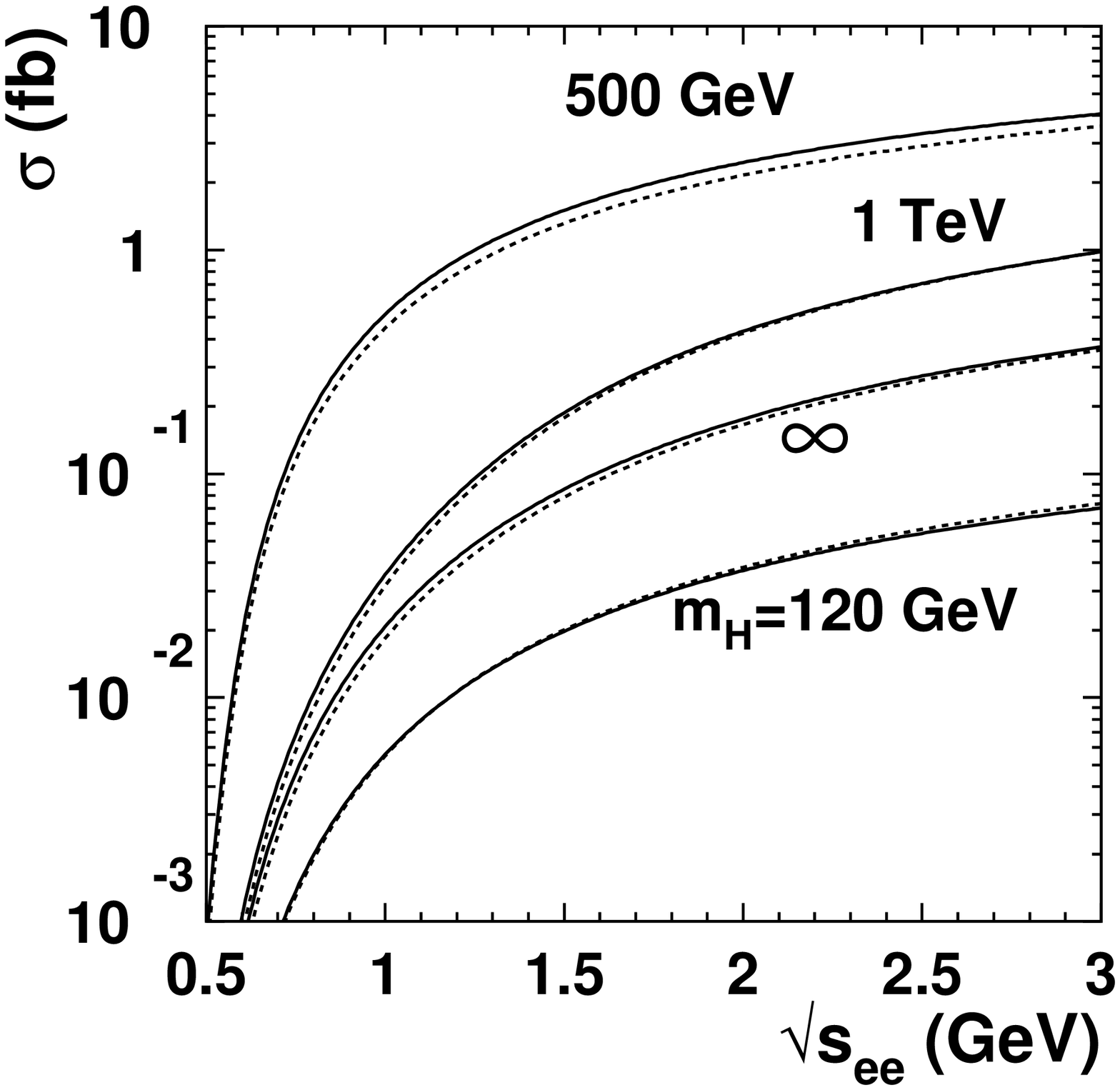,width=2.5in,clip=}}
\caption[]{
Cross sections as a function of $\sqrt{s}_{e^+e^-}$
for (a) $e^+e^- \to \nu\bar{\nu} t \bar t$ via $W^+_LW^-_L$ fusion 
and for (b) $e^+e^-\to e^+e^-  t \bar t$ via $Z_LZ_L$ fusion.
In both cases 
the solid line is the ${\cal O} (\alpha_s) $ 
QCD corrected cross section and the dashed line is the 
electroweak tree level
cross section. }
\end{figure}

In Fig. 5 we show the tree level electroweak and ${\cal O} (\alpha_s) $ QCD 
corrected cross sections for the subprocesses 
$W^+W^- \to t\bar{t}$ and $ZZ \to t\bar{t}$ for a Higgs boson mass 
of 500~GeV and for all $W$ and $Z$ polarizations.  
In all cases we include the $t\bar{t}+g$ final state in the 
${\cal O} (\alpha_s) $ results. 
To separate out the hard gluon case depends on details of
the detector and 
the jet finding algorithms which is beyond the scope of this 
paper.  As before, we include in these and all subsequent results
the kinematic cuts
$m_{t\bar{t}}>400$~GeV and $p_T^{t,\bar{t}}>10$~GeV.  The Born results 
agree with previous results \cite{Gintner:1996cr}.
The longitudinal scattering cross section is much larger 
than the $TT$ and $TL$ cases.  Since it is the longitudinal gauge 
boson processes which corresponds to the Goldstone bosons of the 
theory we will henceforth only include results for $V_L V_L$ scattering. 
In Fig.~6 we show the 
cross section only including longitudinal $W$ and $Z$ 
scattering as a function 
of the $e^+e^-$ centre of mass energy for several representative Higgs 
masses including the $M_H\to \infty$ case (corresponding to the LET).

The QCD corrections to longitudinal scattering 
are often presented 
as a K-factor, normally defined as the ratio of the NLO to LO cross sections. 
Because the ${\cal O} (\alpha_s) $ 
QCD corrections we have calculated in this paper are LO 
corrections to a tree level electroweak result we are taking the 
K-factor factor to be the ratio of the cross section with the 
${\cal O} (\alpha_s) $ QCD 
corrections and the tree level electroweak cross sections. 
The K-factors for 
$\sigma(e^+e^- \to \nu\bar{\nu} t \bar t)$ which goes via
$W^+_LW^-_L$ fusion and for 
$\sigma(e^+e^-\to e^+e^-  t \bar t)$ which goes via $Z_LZ_L$ fusion
are shown in Fig.~7 as a function of the $e^+e^-$ centre of mass 
energy.  K-factors are shown for $M_H=120$~GeV, 500~GeV, 1~TeV, and 
the LET case (using the same kinematic cuts as before).
The ${\cal O} (\alpha_s) $ 
QCD corrections are largest for $M_H=500$~GeV with K-factors  
ranging from over 1.2 for $\sqrt{s}_{e^+e^-}=500$~GeV to 1.15 for 
$\sqrt{s}_{e^+e^-}=1$~TeV. The corrections decrease
as the $e^+e^-$ centre-of-mass energy increases.
The corrections are smallest for a light 
Higgs but in that case we will probably study top-Higgs couplings in 
other processes such as associated top-Higgs production.  As can be 
seen from Fig. 7 the $M_H=1$~TeV and LET cases lie between these two 
extremes.  The variation of the K-factor with $M_H$ is shown more 
explicitly in Fig.~8.  The fact that the K-factor is largest for 
$M_H=5 00$~GeV in Fig.~7 and that it peaks at $M_H\simeq 400$~GeV in Fig.~8 
is a threshold effect which
is an artifact of the kinematic cut we imposed on the $t\bar{t}$ 
invariant mass.  When we pass through the kinematic threshold the 
dominant contribution to the $t\bar{t}$ threshold comes from the 
s-channel Higgs resonance.  This threshold behavior is also seen in 
Fig.~3 of Ref.~\cite{Gintner:1996cr} but at a different value of $M_H$ 
reflecting the different $M_{t\bar{t}}^{cut}$ used in that paper. 
This threshold region also correponds to a region of low $Q^2$ 
relevant to soft gluon emission resulting in larger QCD corrections.  
As the phase space opens up the QCD corrections are expected to 
decrease which is what is observed.
One also sees in Fig.~8
how the K-factor decreases as $\sqrt{s}_{e^+e^-}$ increases.  
The important point one comes away with 
from Fig.~7 and 8 is that it is clear that QCD corrections are not 
insignificant 
compared to the effects we might wish to study such as top Yukawa 
couplings or anomalous $VVt\bar{t}$ couplings.

A brief comment about uncertainties due to scale dependence is in 
order.  As stated above, the scale dependence in this calculation 
only enters in $\alpha_s$ which we took be $\hat{s}$. 
To estimate uncertainties due to scale 
dependence we vary the scale by a factor of two smaller and a
factor of two larger at 
the peak in the subprocess, $\hat{\sigma}$, which is the dominant 
contribution to the $e^+e^-$ cross sections. 
We find an uncertainly of $\sim 15\% $ in $\alpha_s$ which leads to at 
most $\sim 4\%$ uncertainty in the K-factor.

\begin{figure}[t]
\centerline{\epsfig{file=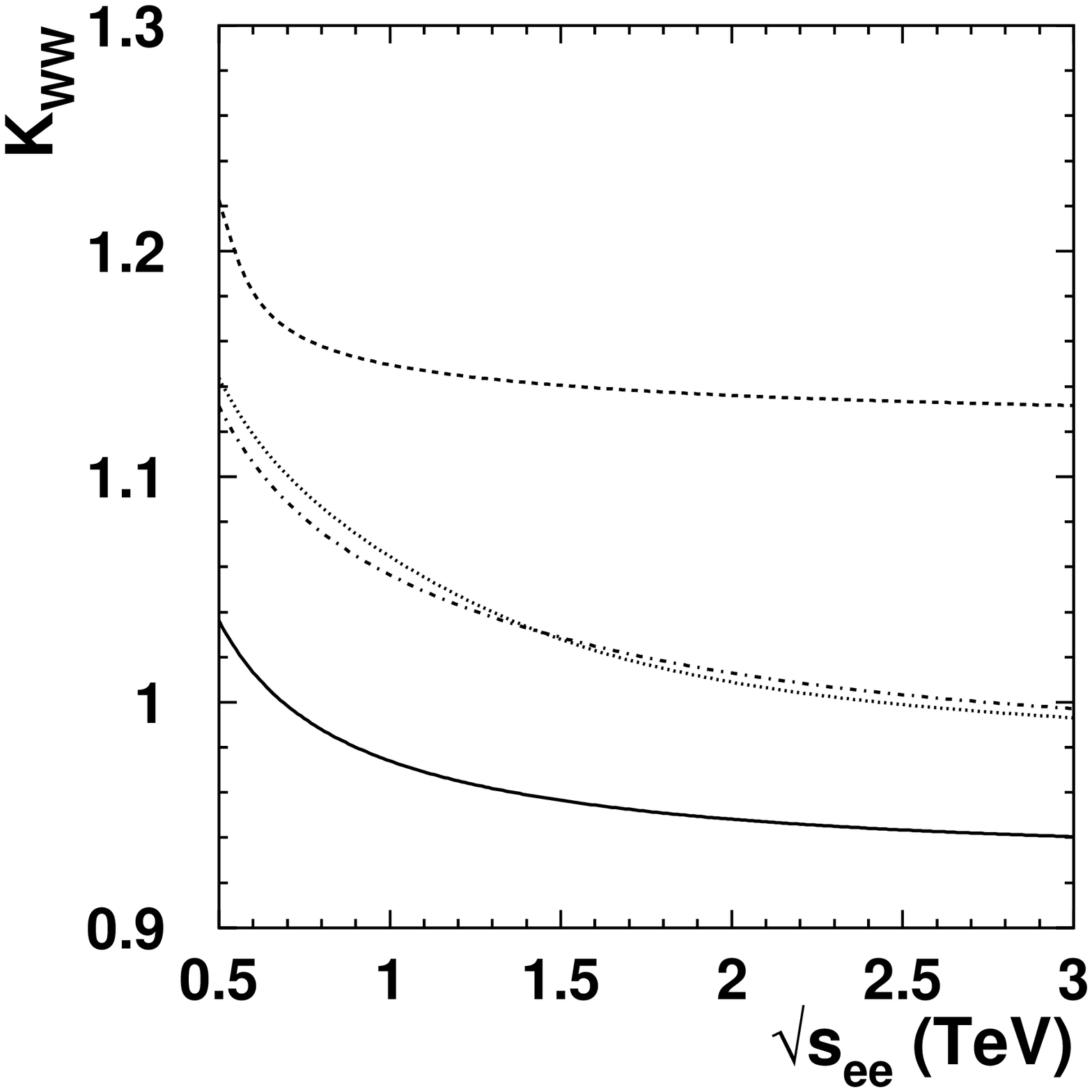,width=2.5in,clip=}}
\centerline{\epsfig{file=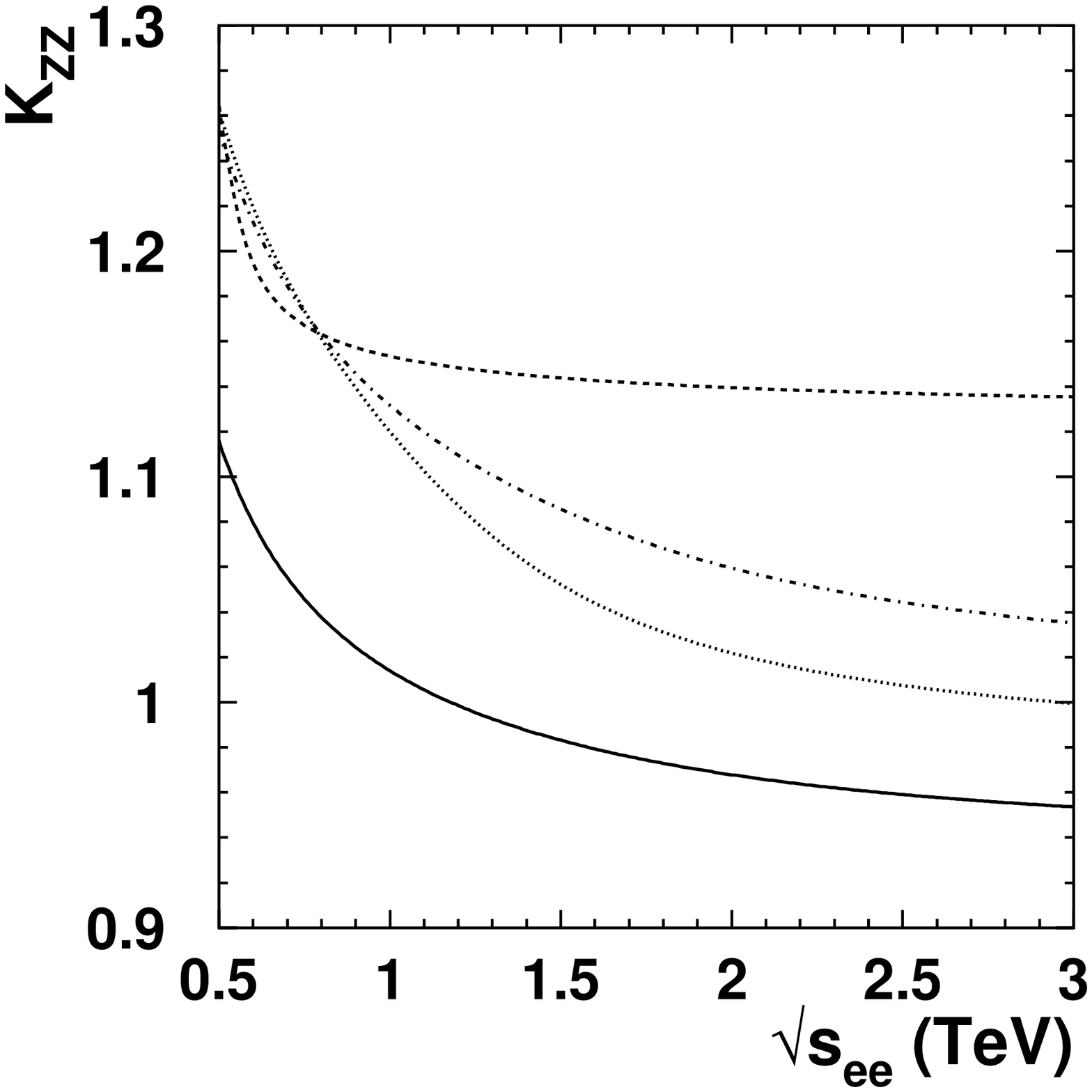,width=2.5in,clip=}}
\caption[]{
The K-factors as a function of  $\sqrt{s}_{e^+e^-}$
for (a) $e^+e^- \to \nu\bar{\nu} t \bar t$ (via $W^+_LW^-_L$ fusion)
and for (b) $e^+e^-\to e^+e^-  t \bar t$ (via $Z_LZ_L$ fusion).
In both cases the solid line is for $M_H=120$~GeV, the dashed line for
$M_H=500$~GeV, the dotted line for $M_H=1$~TeV, and the dot-dashed 
line for $M_H=\infty$ (LET). See text for an explanation of the 
K-factor.
} 
\end{figure}

\begin{figure}[h]
\centerline{\epsfig{file=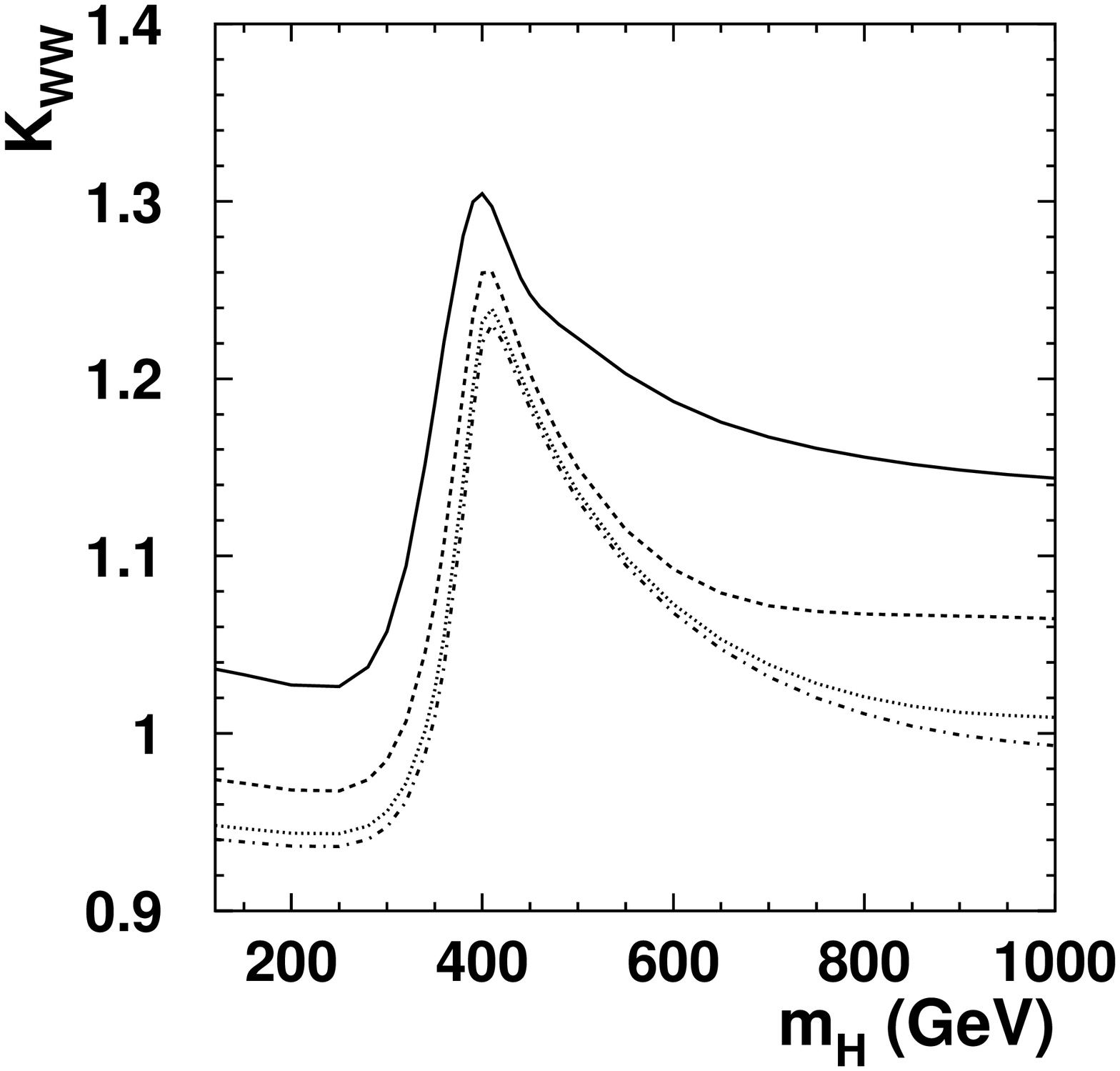,width=2.5in,clip=}}
\centerline{\epsfig{file=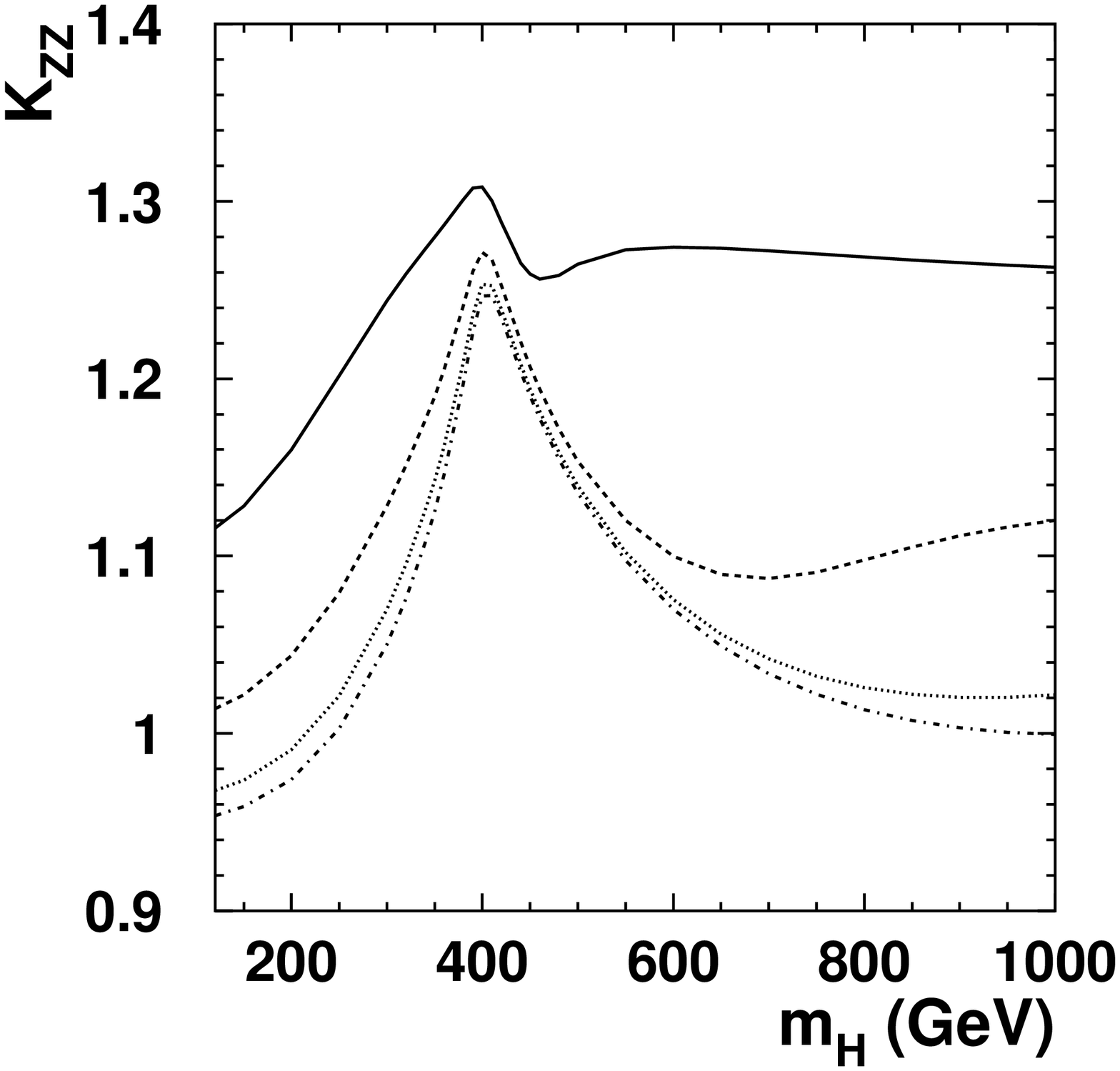,width=2.5in,clip=}}
\caption[]{
The K-factor as a function of  $M_H$
for (a) $e^+e^- \to \nu\bar{\nu} t \bar t$ (via $W^+_LW^-_L$ fusion) 
and for (b) $e^+e^-\to e^+e^-  t \bar t$ (via $Z_LZ_L$ fusion).
In both cases the solid line is for $\sqrt{s}_{e^+e^-}=500$~GeV,
the dashed line for $\sqrt{s}_{e^+e^-}=1$~TeV,
the dotted line for $\sqrt{s}_{e^+e^-}=2$~TeV, and
the dot-dashed line for $\sqrt{s}_{e^+e^-}=3$~TeV.
See text for an explanation of the K-factor.
} 
\end{figure}

\section{Discussion}

In this paper we concentrated on ${\cal O} (\alpha_s) $ QCD corrections.  
There are equally 
important considerations and contributions that we have neglected.  
For completeness we briefly describe them here.  

Backgrounds are always an important ingredient and have been discussed 
in Ref. \cite{Alcaraz:2000xr,Barklow:1997nf,Han:2000ic}.  Briefly, 
the relevant signal is a $t\bar{t}$ pair with missing transverse 
momentum carried by the neutrinos or from the beam electrons not  
observed by the detector.
The largest background is direct $t\bar{t}$ production.  It can be 
suppressed by requiring missing mass greater than some minimum value. 
This background can be further reduced by choosing the jet association 
that best reconstructs the $t$ and $W$ masses.   Another significant
background is $t\bar{t}$ production via other gauge 
boson fusion with the largest being $\gamma\gamma\to t\bar{t}$ 
\cite{Kauffman:1989aq}.  This can be reduced by requiring the missing 
transverse energy to be greater than some value, with a value of 
50~GeV used by Alcaraz and Morales for a TeV collider 
\cite{Alcaraz:2000xr}. The final background we mention is
$e^+e^-\to \gamma t\bar{t}$ where the photon escapes the detector but 
is sufficiently hard to generate the required $p_T$.   Most of this 
background can be eliminated with an appropriate cut on the 
$p_T^{t\bar{t}}$ of the $t\bar{t}$ pair \cite{Larios:1997ey}.  
Additionally the signal could be enhanced using polarized beams 
\cite{Larios:1997ey}.  

At present the electroweak radiative corrections 
for top quark production have not 
been calculated and there are ambiguities which still need to be 
better understood.  In general,
when the Higgs boson mass is heavy, 
the electro-weak correction to the process contains
the enhancement factor $\alpha m_H^2/M_W^2$ which will be significant. 
However,
S-matrix elements of $VV \rightarrow t \bar t$ are not well-defined
for unstable external particles so are problematic
for higher-order calculations that include the Higgs width at the born level.
The discussion of this issue has been given for the process $W^+W^-
\rightarrow W^+W^-$ in \cite{Denner:1997kq}.
The full process with stable initial and final particles needs
to be considered if one includes electro-weak corrections 
consistently. This is beyond the scope of this paper.

The accuracy of the effective $W$ approximation has been studied by a 
number of authors.  Assuming a cut on the invariant mass of 
$M_{t\bar{t}}>500$~GeV the resulting cross sections typically agree to 
about 10\% with the exact calculation \cite{Larios:1997ey}.  
This is the same order of 
magnitude as the effects we are interested in measuring and 
the QCD corrections considered in this paper. Therefore, an exact calculation 
would be required for precision measurements of $t\bar{t}$ cross sections.

Finally, we mention that 
the subprocess $W^+W^-\to t\bar{t}$ is also possible in $\gamma\gamma$ 
collisions but the standard model direct processes overwhelm this 
subprocess by several orders of magnitude.
Analogous to the $e^+e^-$ case one can use the effective $W$ 
luminosity inside photons   \cite{Cheung:1994sa,Hagiwara:1991zz}.
It is possible that 
judicious choices of kinematic cuts could enhance the signal; in 
$\gamma\gamma\to t\bar{t}$ the cross section is dominated by the 
top-quarks collinear to the beam while in the $W$-fusion process 
the spectator $W$'s along the beam direction could be used to tag events
\cite{Cheung:1994sa}.
We leave this process  for a future study.

\section{Conclusions}

In the event that the Higgs mass is heavy and the electroweak sector 
is strongly interacting it is quite possible that the underlying 
theory will manifest itself in the interactions between the top quark 
and longitudinal component of gauge bosons. Different aspects of this 
have been studied but the common theme is that precision measurements 
at a future high energy $e^+e^-$ would be necessary to understand the 
underlying dynamics.  In this paper we studied the 
${\cal O} (\alpha_s) $ QCD corrections to 
the tree level electroweak process $VV\to t\bar{t}$.  
We found that they can be quite substantial, 
the same size as the effects we wish to study, so that they 
need to be taken into account when studying $t\bar{t}$ 
production.  Although the kinematic cuts that are chosen will 
change the numerical results slightly, 
the ${\cal O} (\alpha_s) $ QCD corrections are understood 
and should not pose a barrier in studying these processes.

\acknowledgments
SG thanks Frank Petriello for helpful discussions.
This research was supported in part by
the Natural Sciences and Engineering Research Council of Canada. 
SZ was also supported in part by the Natural Science Foundation of 
China under grant no. 90403004


\end{document}